\documentclass[12pt]{iopart}

\usepackage{iopams}  
\usepackage{amssymb,graphicx}
\usepackage{multirow} 

\begin{document}

\title[Controlling the SPP excitation efficiency using dielectric magneto-optical cavity]{Controlling the surface plasmon excitation efficiency using dielectric magneto-optical cavity} 



\author{D A Bykov and L L Doskolovich}\ead{bykovd@gmail.com}

\address{Image Processing Systems Institute of the RAS, 151 Molodogvardeiskaya st., Samara 443001, Russia}
\address{Samara State Aerospace University, 34 Moskovskoye shosse, Samara 443086, Russia}

\begin{abstract}
The diffraction of a plane wave by a magneto-optical cavity located on a metal interface is investigated theoretically. We show that the excitation of localized eigenmodes of the cavity allows one to efficiently excite the surface plasmon polariton (SPP). We exploit the polarization and symmetry properties of the cavity modes to propose an efficient approach for controlling the SPP intensity through an external magnetic field. The presented theoretical predictions are in good agreement with the rigorous computations based on the generalized Lorentz reciprocity theorem and aperiodic Fourier modal method. The magnetization-induced relative intensity variation of the excited SPP demonstrated in numerical computations varies from a few percent to 100\% depending on the polarization of the incident wave. Such large modulation opens new possibilities for efficient SPP modulation with considered magneto-optical intensity effect.
\end{abstract}


\submitto{\JO}

\maketitle 


\section{Introduction}
 In recent years, considerable attention has been given to the investigation of nanoscale structures for the excitation and manipulation of surface plasmon polaritons (SPP). In a number of papers, a variety of geometries for the highly efficient excitation of the SPP~\cite{r1, r3a} and SPP steering~\cite{r3a, r2,r3, r3b} were proposed and analyzed. Simple theoretical models~\cite{r4} and efficient numerical methods~\cite{r1} to design the corresponding structures have been proposed. 

The hybrid plasmonic structures containing magneto-optical materials open new possibilities for controlling the SPP characteristics on sub-nanosecond time scales. 
In particular, a variety of magneto-plasmonic structures to control the SPP dispersion through an external magnetic field were proposed and investigated~\cite{r5,r6,fedyanin, Khanikaev, Chetvertukhin, belotelov2009josab}. 
In this work we, for the first time, look into the possibility of controlling the SPP's intensity by means of optical cavities made of magneto-optical materials (Fig.~1). 
In our approach the same optical cavity is used both to excite the SPP and to control the excitation efficiency. 
The proposed approach can be used for designing various active plasmonic elements aimed at ultrafast SPP modulation.

\begin{figure}[h]
\centerline{\includegraphics{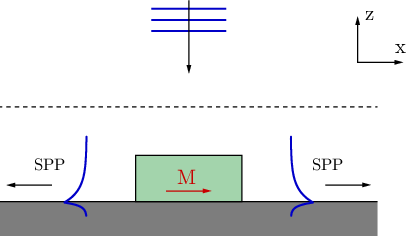}}
\caption{Magneto-optical cavity located on metal interface: the excitation of two SPPs by normally incident plane wave (cavity height is $h=1040\; {\rm nm}$, cavity width is $w=1900\; {\rm nm}$, the cavity is infinite along $y$- direction).}
\end{figure}

The paper is organized in five sections.
Following the Introduction, Section II presents theoretical model of the SPP excitation using magneto-optical cavity. The model is based on analysis of symmetry properties of magneto-optical cavity in 2D case.
In Section III we describe a rigorous numerical method used to calculate the SPP excitation efficiency.
Section IV presents the numerical study of the SPP excitation with a rectangular cavity.
Resonant magneto-optically-controlled SPP excitation is analyzed using coupled-mode theory and scattering matrix formalism.


\section{Theory}
\subsection{Statement of the problem}

Let us consider the problem of the excitation of an SPP using a plane wave normally incident on a cavity located on a metal surface (Fig.~1). 
Denote the amplitude of the SPP excited by non-magnetized cavity as $A(0)$.
Now let us consider magnetized cavity with the magnetization vector $\vec M$ being parallel to the Ox-axis (Fig.~1). 
The optical properties of magnetized cavity are described by the following dielectric permittivity tensor:
\begin{equation}
\label{eq1}
\boldsymbol\varepsilon = 
\left[ \begin{array}{ccc}
\varepsilon & 0 & 0 \\
0& \varepsilon & \mathrm{i} g \\
0 & -\mathrm{i} g& \varepsilon  \\
\end{array} \right] ,
\end{equation}
where $g$ is the absolute value of the gyration vector proportional to applied magnetic field.
Thus, we can denote the amplitude of the SPP excited by magneto-optical cavity as $A(g)$.
Accordingly, the intensity (or excitation efficiency) of the SPP is $|A(g)|^2$.
Now let us introduce the modulation efficiency of the SPP as relative intensity variation due to magnetization:
\begin{equation}
\label{eq4}
\delta = \frac{|A(g)|^2-|A(0)|^2}{|A(g)|^2} \times 100\%.
\end{equation}
To control (or modulate) the SPP intensity it is important to simultaneously obtain high values of $|A(g)|^2$ and of $\delta$.
As we show below, high SPP intensity can be obtained due to the excitation of the cavity modes, 
while high modulation efficiency can be achieved by designing the polarization and symmetry properties of the eigenmodes.


\subsection{Cavity modes symmetry}
Note that in the case of non-magnetic materials the problem in question is symmetrical, with both the incident wave and the cavity showing symmetry. Because of this, as a starting point of our analysis, we need to study the symmetry properties of the cavity modes. 
It can be easily shown that eigenmodes of non-magnetized symmetric structure are either even or odd~\cite{r7,r8}.
Hereafter, we consider the symmetry of the y-component of the field, i.e.\ we say that TE (TM) mode is even (or symmetric) if $E_y(x)=E_y(-x)$ ($H_y(x)=H_y(-x)$). Correspondingly, TE (TM) mode is odd (or antisymmetric) if $E_y(x)=-E_y(-x)$ ($H_y(x)=-H_y(-x)$). Note that the odd modes are unable to be excited by a (\emph{symmetric}) normally incident plane wave.

In this work, we study magnetized structures, with the magnetization vector being perpendicular to the structure's symmetry plane (see Fig.~1). 
Rather than breaking the field symmetry, the said magnetization direction just modifies it as follows~\cite{r7,r8}. 
Following the magnetization, the odd TE-modes of the non-magnetized structure retain odd TE-components, while acquiring even TM-components (Fig.~2). 
Correspondingly, the odd TM-modes retain odd TM-components while acquiring even TE-components.
These statements remain to be valid if the mode symmetry is reversed (if we interchange the words ``odd'' and ``even'')~\cite{r7}. 
The modes that can be excited in the structure are defined by the polarization of the incident wave and the mode symmetry.
In particular, only modes with even TE(TM)-components can be excited by normally incident plane TE(TM)-wave~\cite{r7,r8}.

\begin{figure}[th]
\centerline{\includegraphics{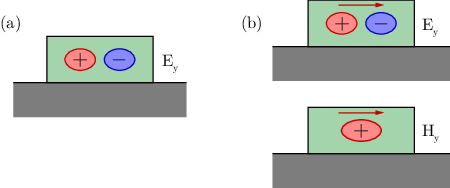}}
\caption{Mode symmetry: field distribution of (a)~odd TE-mode of a non-magnetized structure, (b)~the same mode 
under magnetization: odd TE-component ($E_y$) and even TM-component ($H_y$).}
\end{figure}


\subsection{SPP excitation in non-magnetized structure}
Let us now analyze the SPP excitation by a normally incident plane wave in the case of non-magnetized structure.
Being TM-polarized, the SPP in non-magnetized structure can be excited by TM-wave only.

If there are no eigenmodes supported by the cavity, there occurs the non-resonant scattering of light by the cavity and the excitation of low-amplitude SPPs. If, however, the cavity supports even TM-mode, a resonant scattering of the plane wave from the cavity will take place, which may result, as we show below, in the resonant growth of the SPP intensity. 


\subsection{SPP excitation in magnetized structure}
The magnetization of a structure leads to a change in the cavity mode parameters (symmetry, frequency, quality factor, coupling coefficient), making it possible to control the SPP intensity by an external magnetic field.
For example, the even TM-mode discussed above will change its frequency following the magnetization. 
That, in its turn, will lead to the change of the excitation efficiency $A(g)$.

Moreover, the magnetization gives rise to polarization conversion resulted from non-diagonal terms in permittivity tensor~(\ref{eq1}).
Let us now consider more sophisticated effects governed by ${\rm TE}\leftrightarrow{\rm TM}$ polarization conversion.
Consider a cavity whose eigenmode cannot be excited in the absence of magnetization due to violation of symmetry and/or polarization conditions. For instance, assume that a plane TM-wave strikes a cavity that supports an odd TE-mode. In the absence of magnetization, the said mode is unable to be excited due to the polarization mismatch. However, in that case, there occurs a \emph{non-resonant} excitation of a low-amplitude SPP (see Fig.~3a). 
If the cavity gets magnetized, the eigenmode acquires an even TM-component and will be able to be excited~\cite{r7,r8}, that will cause the resonant scattering of light and, consequently, the excitation of a resonance-enhanced large-amplitude SPP should be expected (Fig.~3b).

\begin{figure}[ht]
\centerline{\includegraphics{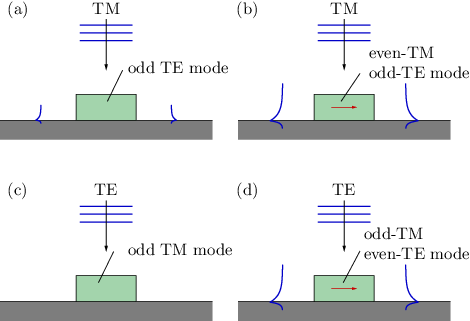}}
\caption{(Color online) Excitation of an SPP by a normally incident plane wave with
(a)~TM polarization in a non-magnetized structure;
(b)~TM polarization in a magnetized structure;
(c)~TE polarization in a non-magnetized structure; 
(d)~TE polarization in a magnetized structure.}
\end{figure}

Of greater interest is the situation when the cavity supporting odd TM-eigenmode is illuminated by a plane TE-wave. In this case, the non-magnetized cavity will be unable to excite an SPP, i.e.\ the intensity of the excited plasmon will be strictly zero (Fig.~3c). If, however, the structure is magnetized, the mode will acquire even TE-components, and therefore can be excited by a TE-wave, thus providing a resonance-enhanced SPP excitation (Fig.~3d). In this case, the magnetization/demagnetization of the structure enables the SPP to be entirely ``switched on/off''.

Similar analysis can be carried if even TE-mode is excited by the TE-wave.
In this case the mode can be excited in non-magnetized structure, however no SPP excitation will take place.
If the structure gets magnetized the eigenmode will acquire TM-components, hence it can be coupled to the SPP-wave, which will result in resonant SPP excitation.

In total, there are eight cases corresponding to different mode symmetry and polarization combinations. 
However, only in four cases presented in Table~1 the mode excitation takes place. 
Note that, due to the symmetry, the cavity always excites two SPPs (the left-propagating and the right-propagating) with equal intensity. However, the SPPs excited by TM-wave (cases~1 and~3) are in phase, whereas the SPPs excited by TE-wave (cases~2 and~4) are out of phase.

\begin{table}[tb]
  \caption{SPP excitation using eigenmodes of different polarization and symmetry}
\lineup
	\begin{tabular}{@{}cccc@{}ccc@{}}
    \br
    \multirow{2}{*}{Case} & \multirow{2}{*}{\parbox{1.4cm}{\centering Incident\\wave}} & \multicolumn{2}{c}{Non-magnetized structure} && \multicolumn{2}{c}{Magnetized structure} \\ 
		\cline{3-4}
		\cline{6-7}
		&& Eigenmode & SPP excitation && Eigenmode & SPP excitation\\
		\mr
    1 & TM & \parbox[c]{2cm}{\centering odd TE: \\not excited} & non-resonant  &&  \parbox[c]{3.6cm}{\centering odd-TE -- even-TM: \\excited} \vspace{0.7em} & resonant\\ 
		2 & TE & \parbox[c]{2cm}{\centering odd TM: \\not excited} & no excitation &&  \parbox[c]{3.6cm}{\centering odd-TM -- even-TE: \\excited} \vspace{0.7em} & resonant\\
		3 & TM & \parbox[c]{2cm}{\centering even TM: \\excited}    & resonant      &&  \parbox[c]{3.6cm}{\centering even-TM -- odd-TE: \\excited} \vspace{0.7em} & resonant\\
		4 & TE & \parbox[c]{2cm}{\centering even TE: \\excited}    & no excitation &&  \parbox[c]{3.6cm}{\centering even-TE -- odd-TM: \\excited} & resonant\\
		\br
  \end{tabular}
\end{table}


\section{Methods}

For a proof-of-principle of this approach we considered the simplest cavity geometry: a rectangular block located on metal interface (geometrical parameters are presented in the caption to Fig.~1). In the course of calculations, we considered silver substrate with its permittivity described by a Drude--Lorentz model~\cite{r9}. The permittivity of the magnetized material was described by a tensor~(\ref{eq1})
with $\varepsilon = 5.06+4.3\times 10^{-4} \mathrm{i}$, $g = 0.015-3\times 10^{-5} \mathrm{i}$ at wavelength $\lambda = 1200\; {\rm nm}$. The said parameters are characteristic to the material $\mathrm{Bi}_{2.2}\mathrm{Dy}_{0.8}\mathrm{Fe}_5\mathrm{O}_{12}$~\cite{r10}.

The SPP's excitation efficiency was estimated using a modification of approach proposed in Ref.~\cite{r1}. 
Instead of solving the `direct' problem of diffraction of a plane wave and calculating the excited SPP's intensity, the `inverse' problem was solved, with the surface plasmon polariton considered to be an incident wave and the plane wave treated as a scattered wave. 

For reciprocal (nonmagnetic) materials, the scattering coefficient \emph{from} the SPP \emph{into} a plane wave (`inverse' problem) is proportional to the excitation coefficient \emph{of} the SPP \emph{by} the plane wave (direct problem). Strictly speaking, this fact is described by the Lorentz reciprocity theorem, which allows one to derive the following relationship for the complex amplitude of a SPP excited by a normally incident TM-polarized plane wave~\cite{r1}:
\begin{equation}
\label{eq2}
A^{\mathrm{TM}} = 
K(\lambda) \int_{-\infty}^{+\infty}H_y(x,z_0) \;\mathrm{d}x,
\end{equation}
where $H_y(x,z)$ is the magnetic field distribution derived from the solution of the `inverse' problem of diffraction of the SPP by a cavity, $K(\lambda)$ is the normalization coefficient defined by the SPP field distribution at a given wavelength~\cite{r1}. The integral in Eq.~(\ref{eq2}) is taken along a straight line $z=z_0$ marked by a dashed line in Fig.~1, with $z_0$ assumed to be sufficiently large~\cite{r1}. Note that the integral in Eq.~(\ref{eq2}) defines the scattering coefficient from the SPP to a plane wave propagating along the Oz-axis. The `inverse' problem was solved using Aperiodic Fourier Modal Method~\cite{r11}.

Because the structure under analysis contains nonreciprocal (magneto-optical) materials, a generalized Lorentz reciprocity theorem needs to be used~\cite{reciprocity}. 
This means that the `inverse' problem should be solved for a modified structure described by a transposed permittivity tensor. For the magneto-optical materials, the transposition of the tensor~(\ref{eq1}) means that the structure is reversely magnetized. 
Taking this into account, Eq.~(\ref{eq2}) remains valid, describing the SPP's excitation efficiency for a normally incident plane TM-wave. As we indicated above, in a magnetized structure, the SPP can be excited using not only a TM-wave but also a TE-wave. In the latter case, the complex amplitude of the SPP excited by the plane TE-wave is given by
\begin{equation}
\label{eq3}
A^{\mathrm{TE}} = 
K(\lambda)\cdot n \int_{-\infty}^{+\infty}E_y(x,z_0) \;\mathrm{d}x,
\end{equation}
where $n=1$ is the surrounding medium refractive index and $E_y(x,z)$ is the electric field distribution derived by solving the `inverse' problem of diffraction of the SPP by a cavity (rectangular block) magnetized in the opposite direction to the Ox-axis.

Note that Eqs.~(\ref{eq2}) and~(\ref{eq3}) can be used to calculate the SPP excitation efficiency for both magnetized and non-magnetized cavities supporting eigenmodes of arbitrary polarization. However, for non-magnetized cavity the value of $A^{\mathrm{TE}}$ in Eq.~(\ref{eq3}) is always zero.


\section{Numerical results and discussions}
Shown in Fig.~4a are the intensities of the SPPs excited by normally incident TM- and TE-waves in non-magnetized structure ($g=0$). 
The intensities were calculated numerically using Eqs.~(\ref{eq2}) and~(\ref{eq3}).
As indicated above, the intensity of the excited SPP is strictly zero for the TE-polarized incident wave.
 Fig.~4b presents the magnetization-induced SPP intensity variation for different polarizations of the incident plane wave. The spectra are seen to contain pronounced resonance peaks denoted as A, B, C, D. 
Each resonant peak corresponds to the excitation of the structure eigenmode. 

\begin{figure*}[t]
\centerline{\includegraphics{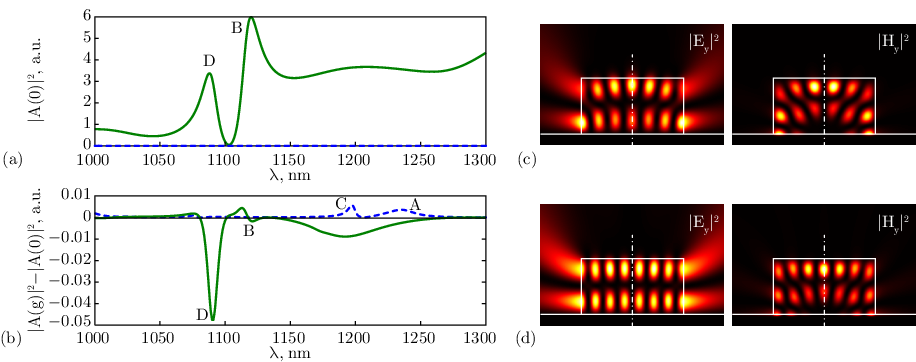}}
\caption{(Color online) (a)~SPP excitation efficiency with a normally incident plane 
TM-wave ($|A^{\mathrm{TM}}|^2$ , green solid curve) 
and 
TE-wave ($|A^{\mathrm{TE}}|^2$ , blue dashed curve) 
in a non-magnetized structure. (b)~Efficiency variation in response to the structure magnetization.
(c)~Field distribution of eigenmode with $\lambda = 1198.4+3.389\mathrm{i}\; {\rm nm}$ (resonance C).
(d)~Field distribution of eigenmode with $\lambda = 1090.2+7.792\mathrm{i}\; {\rm nm}$ (resonance D).}
\end{figure*}

According to temporal coupled-mode theory~\cite{coupledmode} each mode/resonance can be described by the following parameters:
complex wavelength $\lambda_p$,
non-resonant scattering coefficient $a$, 
and two coupling coefficients: from the plane wave to the cavity ($c_1$) and from the cavity to the SPP ($c_2$).
The numerical values of these parameters define the SPP excitation efficiency~\cite{coupledmode}: 
\begin{equation}
\label{fano}
A \approx a+\frac{c_1 c_2}{\lambda-\lambda_p}.
\end{equation}
Eq.~(\ref{fano}) describes the Fano line shape for the peaks in Fig.~4a,b.
In the presence of external magnetic field the parameters of the resonance ($\lambda_p$, $a$, $c_1$, $c_2$) should be considered as functions of the gyration $g$.
Accordingly, the magneto-optical properties of the structure (Fig.~4b) are determined by magnetization-induced variation of above-mentioned parameters of the resonance.


To study the structure eigenmodes we used the scattering matrix approach~\cite{gippius, r12}.
Under this approach the complex eigenfrequencies (wavelengths) of the cavity are calculated as the poles of the scattering matrix analytic continuation. Moreover, this approach allows us to calculate the electromagnetic field distribution of the eigenmodes, thus to analyze the mode symmetry.

Let us consider each resonance in details.
The resonance C at a wavelength of $\lambda = 1197\; {\rm nm}$ corresponds to the excitation of a mode with the following complex wavelength: $\lambda_p = 1198.4+3.389\mathrm{i}\; {\rm nm}$. An analysis of the mode's field distribution (Fig.~4c) suggests that this mode is odd TM-mode (no antinodes of $|H_y|^2$ on the symmetry line), whereas mode's TE components are even (two antinodes of $|E_y|^2$ on the symmetry line). In the non-magnetized structure this mode is an odd TM-mode ($\lambda_p = 1198.4+3.378\mathrm{i}\; {\rm nm}$). In full compliance with the above-specified symmetry conditions, the said mode is excited by a TE-wave, leading to a resonance of the $A^{\mathrm{TE}}(g)$ magnitude.

The resonance A at a 1235-nm wavelength also corresponds to the excitation of an even-TE -- odd-TM mode ($\lambda_p = 1237.3+12.558\mathrm{i}\; {\rm nm}$) of the magnetized structure, but as distinct from the previous mode, the mode in question in the non-magnetized structure represents an even TE-mode ($\lambda_p = 1237.3+12.562\mathrm{i}\; {\rm nm}$).

The resonances B and D of magnitude $A^{\mathrm{TM}}$ ($\lambda = 1113\; {\rm nm}$, $\lambda = 1090\; {\rm nm}$ in Fig.~4a,b) are associated with the excitation of eigenmodes with the complex wavelengths $\lambda_p = 1116.1+8.151\mathrm{i}\; {\rm nm}$ and $\lambda_p = 1090.2+7.792\mathrm{i}\; {\rm nm}$ ($\lambda_p = 1116.1+8.142\mathrm{i}\; {\rm nm}$ and $\lambda_p = 1090.3+7.733\mathrm{i}\; {\rm nm}$ in non-magnetized structure). An analysis of the modes field distribution (see Fig.~4d) suggests that while being an even TM-modes in a non-magnetized structure they are converted into an even-TM -- odd-TE modes following the magnetization.

According to Table~1, resonances A, B, C and D correspond to the cases 4, 3, 2 and 3, respectively.
The calculated complex wavelengths of the structure eigenmodes are in good agreement with the spectral positions and FWHM of the intensity peaks~(Fig. 4ab). 
Note that along with the four considered modes the cavity supports a number of modes with lower quality factor. These modes govern broad features in the spectrum.

Thus, according to Fig.~4, in non-magnetized structure the SPP can be excited by TM-polarized wave only. 
In this case, the SPP intensity strongly increases when the localized cavity modes are excited (resonances B, D on Fig.~4a).
Magnetization of the structure allows one to excite the SPP using TE-polarized wave as well.
As a rule, the SPP excitation efficiency in the case of TE polarization is lower that that in case of TM polarization.
Moreover, with the incident TM-wave, the magnetization of the structure makes it possible to achieve a larger value of the SPP intensity modulation (see Fig.~4b).
However, the maximal value of \emph{relative} modulation efficiency~(\ref{eq4}) for TM polarization is $\delta^{\mathrm{TM}}=2\%$. 
At the same time for TE polarization $A^{\mathrm{TE}}(0)=0$ and we obtain $\delta^{\mathrm{TE}}=100\%$.
Thus, in the case of TE-polarized incident wave the excited SPP can be totally ``switched off'' by the cavity demagnetization and ``switched on'' by magnetization.


\section{Conclusion}
In conclusion, we considered the SPP excitation using magneto-optical cavity located on metal interface.
We have shown that the excited SPP intensity can be significantly increased in the case of the cavity mode excitation.
The design of symmetry and polarization properties of magneto-optical cavities allows one to achieve high SPP intensity modulation.
In particular, one can excite the SPP using TE-polarized wave to attain a relative magnitude of SPP intensity modulation equal to 100 percent.


Note that in the current paper we focused on the SPP excitation problem because SPP is the simplest mode that can be considered. 
However, the results obtained are more general: the very same approach can be applied to control the excitation of the guided modes of slab waveguides or photonic wires, thus avoiding the SPP's shortcomings caused by signal attenuation.



%
%

%

\ack
The work was financially supported by 
Russian Foundation for Basic Research (RFBR) grants 12-07-00495, 13-07-00464, 13-07-97001,
by Russian Science Foundation (RSF) grant 14-19-00796,
by the ministry of education and science of the Russian Federation (RF), 
and by RF Presidential scholarship SP-1665.2012.5.

\section*{References}
\bibliographystyle{unsrt}
\bibliography{bykov}

\end{document}